# Tunable fluid-solid metamaterials for manipulation of elastic wave propagation in broad frequency range

Quan Zhang,[1] Kai Zhang,[1,a)] and Gengkai Hu[1]

[1]*School of Aerospace Engineering, Beijing Institute of Technology, Beijing, 100081, China*

a) zhangkai@bit.edu.cn

**Abstract**

Current strategies for designing tunable locally resonant metamaterials are based on tuning the stiffness of the resonator; however, this approach presents a major shortcoming as the effective mass density is constant at high frequency. Here, this paper reports a type of tunable locally elastic metamaterial—called "tunable fluid-solid composite"—inspired by the functions of heart and vessels in animals and humans. The proposed metamaterial consists of several liquid or gas inclusions in a solid matrix, controlled through a pair of embedded pumps. Both the band gap and effective mass density at high frequency can be tuned by controlling the liquid distribution in the unit cell, as demonstrated through a combination of theoretical analysis, numerical simulation, and experimental testing. Finally, we show that the tunable fluid-solid metamaterial can be utilized to manipulate wave propagation over a broad frequency range, providing new avenues for vibration isolation and wave guiding.

Elastic metamaterials have attracted significant interest in recent years because of their broad range of applications, including vibration isolation,[1,2] wave guiding,[3-6] cloaking,[7-11] and focusing.[12-14] Practically, owing to their fixed microstructure, traditional elastic metamaterials only operate over fixed frequency ranges, which limits additional potential applications. To address this issue, electromagnets,[4,15] shape memory effect,[16] structural deformation,[17,18] fluid-structure interaction,[19,20] and piezo-shunting[21,22] have been employed to achieve elastic metamaterials with tunable band gaps. Essentially, tunable locally resonant metamaterials can be divided into two types: those obtained by tuning the stiffness and those realized by tuning the resonator mass. Currently, all strategies for designing tunable locally resonant metamaterials are based on the former approach. Among these, piezo-shunting, which is often used to change the effective stiffness of the resonator, appears as the most promising strategy. However, two challenges still remain. First, the effective mass density of locally resonant metamaterials tuned by piezo-shunting is constant at high frequency,[23] contrary to the requirements of the transformation method based on density regulation. On the other hand, current piezoelectric materials are friable, which imposes only small deformations in the object structure. As a result, tuning method based on piezo-shunting cannot be applied to flexible structures.

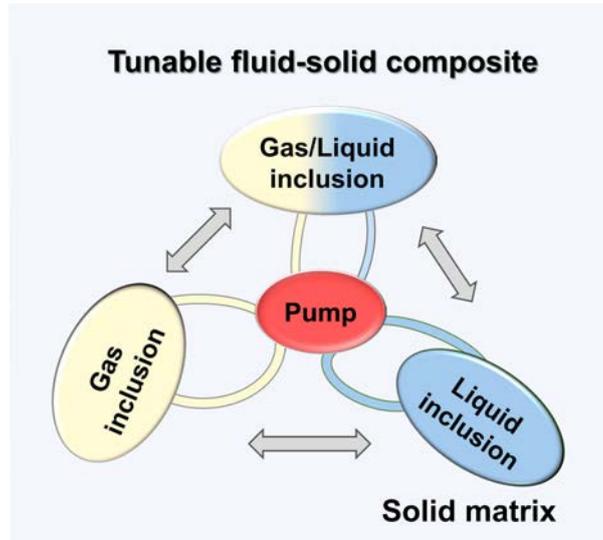

FIG. 1. Proposed tunable fluid-solid composite (tunable multiphase composite) containing several liquid or gas inclusions and a central pump in a solid matrix.

Theoretically, the effective mass density of locally resonant metamaterials at high frequency is simply determined by the substrate mass and can be easily tuned if the mass distribution between resonator and substrate can be modified. However, until now, no studies have investigated the control of mass distribution in the unit cell of elastic metamaterials. The main difficulty is that most elastic metamaterials are solid, resulting in a fixed constant mass distribution upon fabrication. In this work, inspired by the functions of heart and vessels in animals and humans, namely, by how the heart controls the blood circulation in various organs, a new type of tunable elastic metamaterial, called "tunable fluid-solid composite (tunable multiphase composite)", is developed [Fig. 1]. The proposed tunable fluid-solid composite (tunable multiphase composite) is based on a solid matrix containing several liquid or gas inclusions whose distributions can be easily tuned by controlling a central pump. Thus, the composite can be concretely applied to design tunable band gaps and effective mass density by controlling the liquid distribution in the unit cell. This study provides a new perspective for the design of tunable metamaterials based on the concept of tunable fluid-solid composite, which can be applied to vibration isolation, wave guiding, cloaking.

Effective mass density can be used to describe and predict wave propagation in elastic metamaterials.[24] Combined with the transformation method,[23] the propagation of an elastic wave along a designed path can be realized by tailoring the effective mass density, as shown in Fig. 2(a). Considering a typical locally resonant metamaterial [Fig. 2(a)], the effective mass density can be estimated through Lorentz model:[24]

$$\rho_{\text{eff}} = \left[ \frac{(\omega/\omega_0)^2}{1+r-(\omega/\omega_0)^2} \times \frac{1}{1+r} + 1 \right] \bar{\rho}, \quad (1)$$

where $r = m_1/m_2$ ($m_1$ and $m_2$ are the masses of substrate and resonator, respectively), $\omega_0 = \sqrt{K_0/M_0}$ ($M_0 = m_1 + m_2$ is a constant; $K_0$ is the stiffness of the resonator), $\bar{\rho} = M_0/V$ ($V$ is the effective volume of the unit cell). Figure 2(b) shows the relationships between $\rho_{\text{eff}}/\bar{\rho}$ and $\omega/\omega_0$ for different values of $r$. Notably, by controlling the mass distribution between $m_1$ and $m_2$, the band gap of the lattice structure, which corresponds to the frequency range in which the effective mass density is negative, can be significantly altered. Furthermore, the effective mass density at high frequency varies with $r$. Therefore, controlling the mass distribution between resonator and substrate not only changes the locally resonant band gap, but also allows altering the effective mass density at high frequency, providing a promising approach to design tunable locally resonant metamaterials.

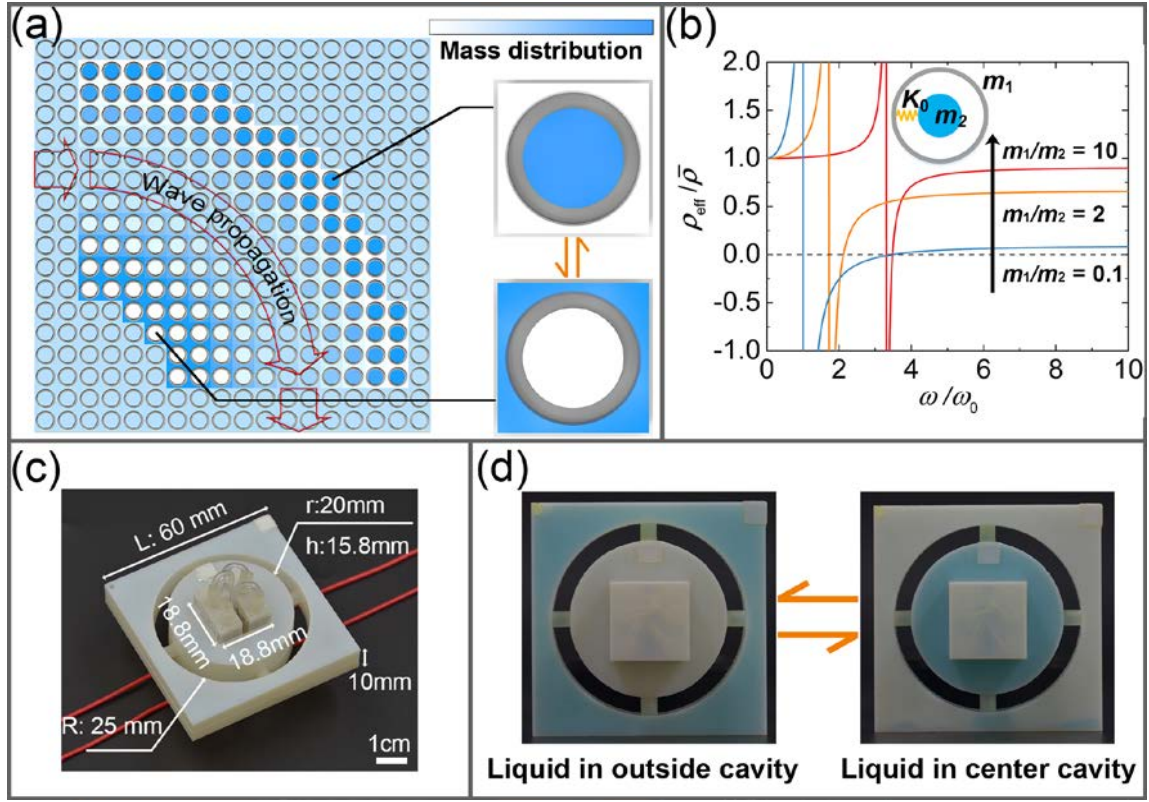

FIG. 2. Designed tunable locally resonant metamaterials. (a) Manipulation of wave propagation along a designed path by controlling mass distribution in different unit cells. (b) Effective mass density at different mass ratios between substrate and resonator. (c) Dimensions of the designed unit cell, scale bar is 1 cm. (d) The liquid is transferred between central and external cavities by the pumps.

To realize this theoretical design, a pair of cavities are initially built by using a three-dimensional (3D) printer (Object 350, Stratasys, USA) with photosensitive resin (RGD835, VeroWhitePlus, Stratasys, USA; Young's modulus $E$ = 2 GPa, which varies little with the increase of frequency at room temperature;[25] density $\rho$ = 1200 kg m$^{-3}$; Poisson's ratio is assumed to be 0.33; Figs. 2(c) and 2(d)). The wall thickness $t$ of the pair of cavities is equal to 0.7 mm, while the other geometrical dimensions are shown in Fig. 2(c). To build the unit cell, the cavities are connected through four tubes (outer diameter: 4.0 mm; inner diameter: 2.0 mm) made of silicon elastomer with Young's modulus $E_r$ = 20 MPa, density $\rho_r$ = 500 kg m$^{-3}$. We prove that the variation of Poisson's ratio of the tube between 0.33 and 0.49 has little effect on the dispersion curves in supplementary material.[26] Finally, two micro pumps, used to transfer the liquid between the two cavities, are glued and embedded into the central cavity. Two tubules connect each pump with the central and external cavities. The total mass of the pumps, attached tubules, and thin lids is 15.72 g. To ensure that the liquid can be completely transferred from one cavity to the other, the volumes of the central and external cavities are designed to be equal. The central cavity, together with the pumps and tubes, acts as a local resonator, while the outside cavity behaves as a substrate. When the mass distribution of liquid in the central and external cavities is tuned, the mass of the local resonator changes. As a consequence, the natural frequency of the resonator will vary, determining the frequency range of the local resonance band gap.

In this work, a numerically based effective medium method[23] is also used to determine the out-of-plane effective mass density of the designed elastic metamaterial. In the numerical simulation, by applying the time-harmonic displacement constraints with $u = 0$, $v = 0$, and $w = Ae^{i\omega t}$ on the external surrounding boundaries of the unit cell, the effective mass density can be expressed as:

$$\rho_{\text{eff}} = -\frac{F_z}{\omega^2 A V_{\text{eff}}}, \qquad (2)$$

where $F_z$ is the amplitude of the effective resultant force, $V_{\text{eff}}$ is the effective volume of the unit cell, and *A* and $\omega$ are the amplitude and angular frequency of the applied displacement, respectively.

Figure 3(a) shows that the out-of-plane effective mass density of the proposed tunable elastic metamaterial can be affected by the liquid distribution in the unit cell. Numerical calculations on the effective mass density of the metamaterial shows very good agreement with the results estimated through Eq. (1). For a given time-harmonic displacement, the interaction force between the resonator and substrate, which contributes to the total effective resultant force ($F_z$), varies with the mass of the resonator. Therefore, controlling the liquid distribution between resonator and substrate through embedded pumps results in different amplitudes of effective resultant force of the unit cell [Eq. (2)], further tailoring the effective mass density of the proposed metamaterial.

To validate the theoretical and numerical predictions, a sample consisting of 10 × 1 unit cells is fabricated. In the experimental setup shown in Fig. 3(c), the plane wave vibrating in the out-of-plane direction is excited by an electrodynamic shaker (HEV-50, Nanjing Foneng, China), which is directly connected to one end of the sample to provide harmonic signal over a broadband frequency range. The dynamic responses at different frequencies are recorded by using two displacement sensors (IL-30, KEYENCE, Japan) attached at both ends of the sample. The transmittance is computed as the ratio between output and input displacement signals, written as $20\log_{10}\|A_{out}(\omega)/A_{in}(\omega)\|$.

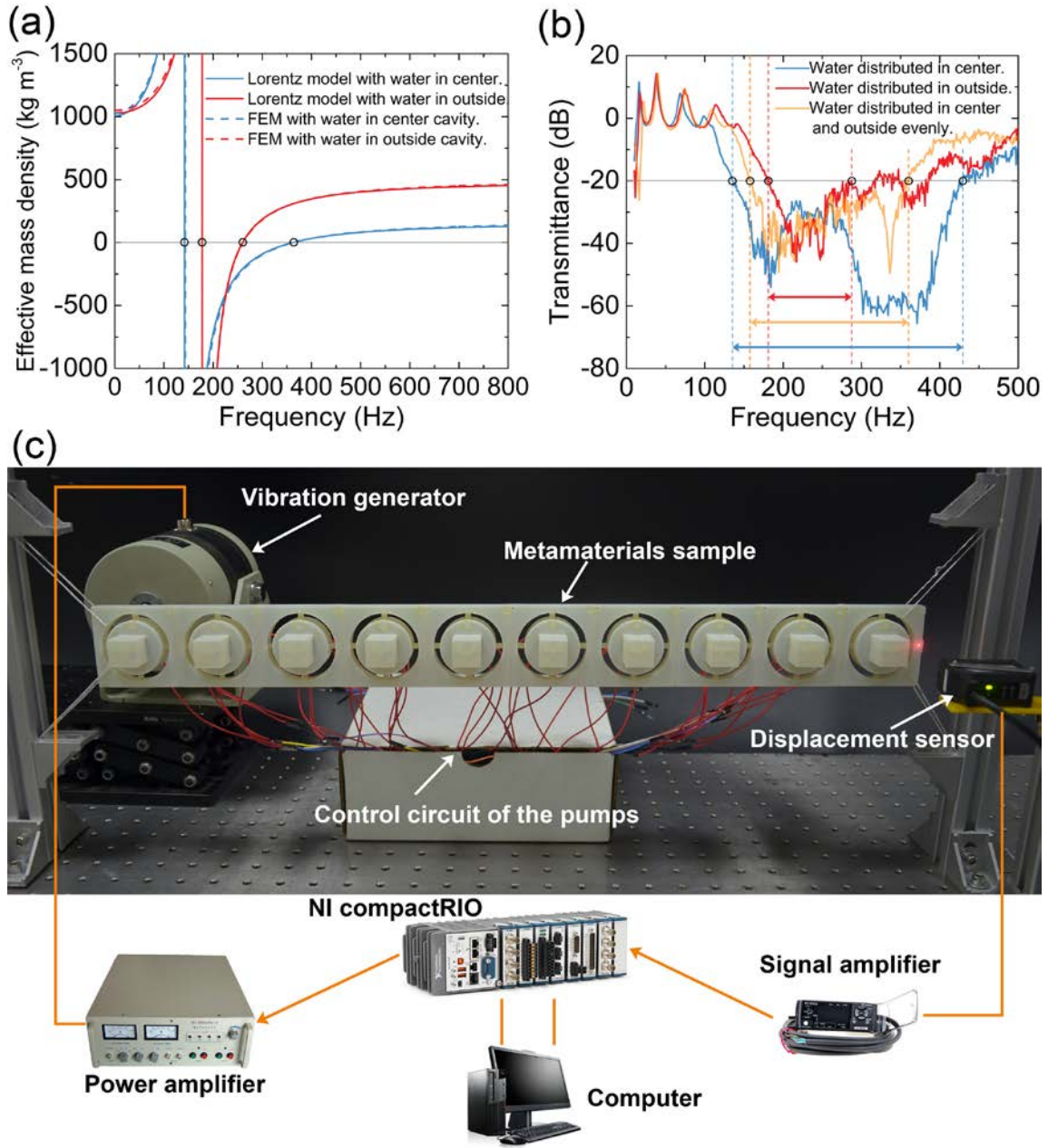

FIG. 3. (a) Calculated out-of-plane effective mass density of the proposed tunable fluid-solid metamaterial. (b) Experimental transmittance curves of the sample comprising 10 × 1 unit cells for different liquid distributions. (c) Experimental setup for the measurement of transmittance curves.

Figure 3(b) shows the experimentally measured transmittance curves for three cases, that are liquid distributed in center cavity, liquid distributed in outside cavity, liquid distributed evenly in center and external cavities, respectively. We also calculate the numerical transmittance curves with the effect of the structural damping coefficient of the elastomer tubes ($\eta$ = 0.4[27-29]), as shown in Fig. S2 in Ref. 26. The overall trend of the measured transmittance curve coincides with that of the calculated one.

In order to quantitatively compare the stopbands on the measured transmittance curves with the predicted band gaps in Fig. 3(a), according to the previous references,[17,19,29-31] the measured bandwidth is defined by selecting a specific attenuation value as a sufficient vibration attenuation threshold. Here the attenuation value is chosen to be 20 dB, which is also used in relevant literatures.[17,29] The defined stop bands are marked by the two-headed arrows in Fig. 3(b). It should be noted that for the case water is distributed in the outside cavity (red curve), the transmittance fluctuates slightly around -20 dB over frequency range from 287 Hz to 385 Hz, we select the first point (287 Hz) reaches -20 dB as the upper boundary of the stop band.

As shown in Fig. 3(b), the lower boundary of the stop band for liquid distributed in the outside cavity is 181 Hz, which matches well with the resonant frequency 177 Hz predicted in Fig. 3(a) (the singular point of the red curve). When liquid is transferred into the center cavity, the lower boundary of stop band defined from the measured transmittance curve changes to be 135 Hz, which also matches well with the predicted resonant frequency 143 Hz in Fig. 3(a). On the other hand, the gap width on the measured transmittance curve for liquid distributed in the outside cavity is 106 Hz, wider than the predicted gap width 84 Hz (the frequency range with negative mass density in Fig. 3(a)). Similarly, the gap width on the measured transmittance curve for liquid distributed in center cavity (294 Hz) is also wider than the predicted gap width (222 Hz). According to the reference[32] and our numerical simulations,[26] the difference is mainly caused by the damping of the elastomer tubes which are used to connect the cavities. The damping of resonator decreases the attenuation over the band gap and increases the attenuation over the frequency range beyond the band gap. Therefore, the experiments indicate that our design is effective.

In the proposed tunable locally resonant metamaterials, liquid distribution not only affects the elastic wave vibrating along the out-of-plane direction, but it also perturbs the elastic wave vibrating along the in-plane direction. To further demonstrate this assertion, we numerically investigate the effect of mass distribution on the propagation of small-amplitude elastic waves through dispersion analysis.[33] Figure 4(a) shows the dispersion curves for the case in which the liquid is completely distributed in the outside cavity. Figure 4(b) shows the Bloch mode shapes of the six lowest bands at the high-symmetry point M, and three typical bands at point G of the IBZ. Notably, all these modes at point M show strong localized vibration, with only the central cavity vibrating while the outside cavity is at rest. Specifically, the first localized mode is associated with the translational vibration in the out-of-plane direction, while the fourth and fifth localized modes are associated with the translational vibration in the in-plane direction. As a result, as shown in Fig. 4(a), these translational vibrations of the local resonator generate band gaps related to propagating waves vibrating in the out-of-plane direction (highlighted in blue, which is consistent with the frequency range with negative mass density in Fig. 3(a)) and in-plane direction (highlighted in red), respectively. Meanwhile, the second, third, and sixth localized modes are associated with the rotational vibrations of the resonator, which do not interact with the propagating waves.[18]

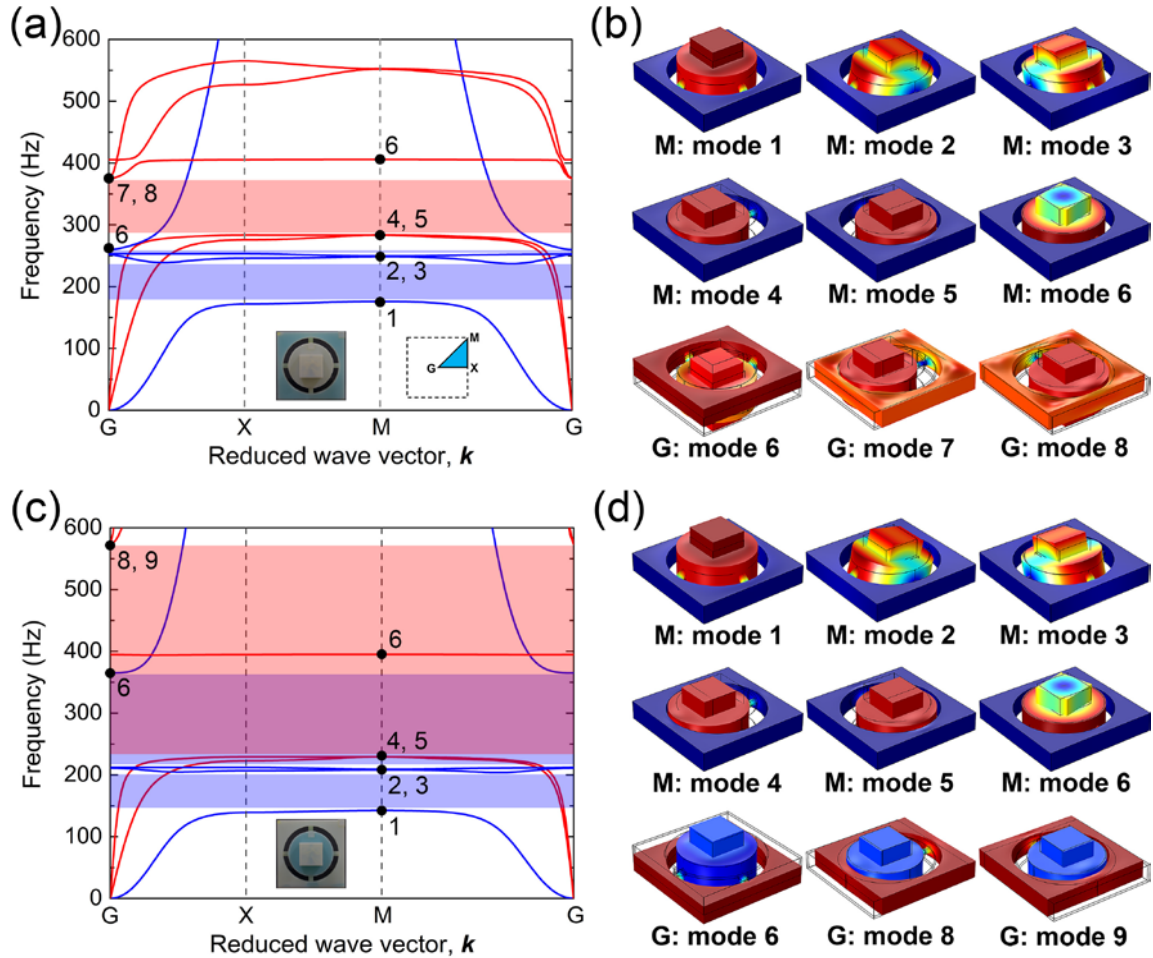

FIG. 4. (a) Dispersion curves and (b) typical Bloch mode shapes at the high-symmetry point for liquid distribution in the outside cavity. (c) Dispersion curves and (d) typical Bloch mode shapes at the high-symmetry point for liquid distribution in the central cavity.

The effect of the mass distribution on the propagation of elastic waves when the liquid is transferred from the external to the central cavity is shown in Figs. 4(c) and 4(d). All the natural frequencies correspond to translational vibration modes of the resonator reduce, as the mass of the resonator increases while the stiffness remains unchanged. Moreover, the frequency ranges of the band gaps become wider as a result of the increase in mass ratio between the resonator and substrate. In addition, there is a complete bandgap for liquid distributed in the center cavity (the overlap region in Fig. 4(c)). Correspondingly, in the measured transmittance curves [Fig. 3(b)], the blue curve shows a strong attenuation as large as 60 dB from 300 to 380 Hz. In contrast, when liquid is transferred into the outside cavity, the complete bandgap disappears, the attenuation reduces from 60 dB to 20 dB over the same frequency range. Notably, the unit cell dispersion analysis is scale invariant,[34] therefore the findings of this work can be applied at different length scales.

The main function of liquid in our design is being controlled to change the mass distribution between the resonator and substrate. Here the nonlinear dynamic behaviors play a weak role in our experiments. In our experiments, for the cases that liquid is distributed in only one cavity, there is no nonlinear dynamic behavior induced by the slosh of liquid since the liquid fully fills the cavity. For the case that liquid is distributed evenly in the center and external cavities, the vibration amplitude is kept to be very small in our experiment. Furthermore, the measured transmittance curve for this case is consistent with the simulated one (where the nonlinear dynamic effect is neglected). For an arbitrary liquid distribution state which is not involved in our experiments, the slosh of liquid can be avoided to a great extend through partitioning the cavity, as illustrated in Ref. 26.

In tuning methods such as piezo-shunting, the frequency range over which the effective mass density meets the requirement of the transformation method is limited, as the effective mass density is constant at high frequency.[23] Conversely, in our proposed tunable fluid-solid composite, the effective mass density at high frequency can be tuned by controlling the mass distribution in the unit cell; this can be utilized to manipulate the wave propagation through a transformation method based on density regulation, as demonstrated in Ref. 26. Compared with the tuning method based on piezo-shunting, the approach proposed here provides several other benefits. First, fluid-solid metamaterials are cheap and easy to fabricate by 3D printing technology; additionally, this method can be applied more suitably to flexible structures, as the cavities can be made of soft materials. Besides, the band gap can be conveniently broadened by choosing different liquid materials; for example, when water is replaced with mercury or a liquid metal, the tuning ranges of the band gap can be significantly broadened.

In summary, we proposed a type of tunable elastic metamaterial—called "tunable fluid-solid composite"—inspired by the functions of the heart and vessels in animals and humans. Both numerical and experimental results indicate that the band gaps of the proposed metamaterials can be tuned by controlling the liquid distribution in the unit cells. Compared with the current strategies based on tuning the stiffness of the resonator, our method provides remarkable advantages as the effective mass density at high frequency can be tuned by controlling the mass distribution. This fluid-solid metamaterial can be applied to flexible structures, providing a better perspective for the design of intelligent systems for vibration isolation, waveguide.

**Acknowledgements**

This work has been supported by the National Natural Science Foundation of China (Grants No. 11672037, Grant No. 11290153 and No. 11632003).